\documentstyle[amstex]{article}

\def\q#1{#1\kern-2pt\char'47}

\font\lowrm=cmr10 scaled \magstep 0

\newcommand{\bquark}{\mbox{$\rm b$}}
\newcommand{\abquark}{\mbox{$\rm \overline{b}$}}
\newcommand{\aNb}{\mbox{$\rm \overline{N_{b}}\ $}}
\newcommand{\Bzero}{\mbox{$\rm B^{0}\ $}}
\newcommand{\Bzerods}{\mbox{$\rm B^{0}_{d,s}\ $}}
\newcommand{\Bplus}{\mbox{$\rm B^{+}\ $}}
\newcommand{\aBzero}{\mbox{$\rm \overline{B}^{0}\ $}}
\newcommand{\Bzerod}{\mbox{$\rm B^{0}_{d}\ $}}
\newcommand{\aBzerod}{\mbox{$\rm \overline{B}^{0}_{d}\ $}}
\newcommand{\Bzeros}{\mbox{$\rm B^{0}_{s}\ $}}
\newcommand{\aBzeros}{\mbox{$\rm \overline{B}^{0}_{s}\ $}}
\newcommand{\Jpsi}{\mbox{$\rm J/\psi\ $}}
\newcommand{\pizero}{\mbox{$\rm \pi^{0}\ $}}
\newcommand{\piplus}{\mbox{$\rm \pi^{+}\ $}}
\newcommand{\pimin}{\mbox{$\rm \pi^{-}\ $}}
\newcommand{\Kzero}{\mbox{$\rm K^{0}\ $}}
\newcommand{\Kzeros}{\mbox{$\rm K^{0*}\ $}}
\newcommand{\photon}{\mbox{$\rm \gamma\ $}}
\newcommand{\muon}{\mbox{$\rm \mu\ $}}
\newcommand{\phph}{\mbox{$\rm 2\gamma\ $}}
\newcommand{\muX}{\mbox{$\rm \muon^{+} X\ $}}
\newcommand{\mumu}{\mbox{$\rm \muon^{+}\muon^{-}\ $}}
\newcommand{\pipi}{\mbox{$\rm \pi^{+}\pi^{-}\ $}}
\newcommand{\bbarb}{\mbox{$\rm \bquark\overline{\bquark}\ $}}
\newcommand{\BJK}{\mbox{$\Bzerod \rightarrow \Jpsi \Kzero\ $}}
\newcommand{\BJKs}{\mbox{$\Bzerod \rightarrow \Jpsi \Kzeros\ $}}
\newcommand{\BJKP}{\mbox{$\Bzerod \rightarrow \Jpsi \Kzero \pizero\ $}}
\newcommand{\KKP}{\mbox{$\Kzeros \rightarrow \Kzero \pizero\ $}}
\newcommand{\PGG}{\mbox{$\pizero \rightarrow \phph\ $}}
\newcommand{\BJKsMMPP}{\mbox{$\Bzerod \rightarrow \Jpsi \Kzero
                      \rightarrow \mumu \pipi\ $}}
\newcommand{\lumin}{\mbox{$10^{33}~cm^{-2}s^{-1}\ $}}

\title{A Possible Measurement of CP-Violation in \Bzero -decays by ATLAS on
LHC}

\author{M\'{a}ria Smi\v{z}ansk\'{a} and Julius H\v{r}ivn\'{a}\v{c}\\
       {\lowrm Fz\'{U} \v{C}SAV, Na Slovance 2, Praha, Czechoslovakia}}

\date{}

\begin{document}


\maketitle

\begin{abstract}

  A possibility to measure CP-violation in \Bzero -decays by ATLAS
experiment was investigated. With one year of running at luminosity
\lumin\\ with a muon $p_{t}$~trigger threshold of 20 $GeV$ and
rapidity coverage $|\eta|<2.5$ of the tracking detector the rate
of 1490 events \BJKsMMPP\\ can be reached.

\end{abstract}

\section*{Introduction}

   The Standard model predicts sizable CP-violating effects from
experimentally established constrains on Kobayashi-Maskawa matrix
elements. The large observable asymmetries are expected in \Bzero and
\aBzero mesons decaying into CP eigenstates \cite{Dib}. Experimentally the
most accessible is a measurement of CP-violation in the
decay  \BJKsMMPP which is convenient
because of low combinatorial background of muons per event.
At LHC the expected production cross section for \bquark\ quarks is large,
$\sigma(\bbarb) \approx 100\mu b \div 700\mu b$ \cite{Aachen}. In order to
avoid
multiple interactions per bunch crossing the luminosity will be taken
\lumin.

\section*{The Method of Measurement}

   The method of measurement of CP-violation will consists in
investigation of number of events $N$ with final state \Jpsi \Kzero as
a function of time-decay of \Bzerod meson. Because of mixing \Bzero ---
\aBzero
the tagging of associated \bquark -particle has to be done by charge of
muon in semileptonic decay \muX. The state \Jpsi \Kzero \muX is
produced by one of the combinations \Bplus \aBzerod, \Bzeros \aBzerod,
\Bzerod \aBzeros, \aNb \aBzerod. The time-dependent rates for
these cases are \cite{Biggi}:

\begin{align*}
  {dN \over dt} &\sim
    {|T_{d}|^{2} \over \Gamma_{+} \Gamma_{d}} e^{-t} (1-n \sin (x_{d}t)) \\
  {dN \over dt} &\sim
    {|T_{d}|^{2} \over \Gamma_{s} \Gamma_{d}} e^{-t} (1-{n \over 1+x^{2}_{s}}
    (x_{s} \cos (x_{d}t) + sin (x_{d}t)) \\
  {dN \over dt} &\sim
    {|T_{d}|^{2} \over \Gamma_{d}^{2}} e^{-t} (1-{n \over 1+x^{2}_{d}}
    (x_{s} \cos (x_{d}t) \pm sin (x_{d}t)) \\
  {dN \over dt} &\sim
    {|T'_{d}|^{2} \over \Gamma_{N} \Gamma_{d}} {e^{-t} (1-n \sin (x_{d}t))}
\end{align*}
$n$ \dots parameter, describing CP-violation effect                   \\
$t$ \dots \aBzerod decay time in lifetime units                       \\
$x_{d}={\Delta m_{d} \over \Gamma_{d}}$ \dots mixing parameter of
                                              \Bzerod \aBzerod mesons  \\
$x_{s}={\Delta m_{s} \over \Gamma_{s}}$ \dots mixing parameter of
                                              \Bzeros \aBzeros mesons  \\
$\Gamma_{+} \Gamma_{d,s} \Gamma_{N}$ \dots decay widths of \Bplus,
  \Bzerods, \aNb respectively

Assuming that the beauty hadron decays essentially due to the decay
of \bquark (\abquark) quark relations holds: $T_{d}=T'_{d}$;
$\Gamma_{+}=\Gamma_{d}=\Gamma_{s}=\Gamma_{N}$.
The observed distribution will be obtained by adding these equations
each of them weighted by it's production value.

   The time-dependent measurement is possible only if the experimental
precision of \Bzerod time-decay reaches the value $\approx \Delta m$,
 which means
a precision of $400 \mu m$ in decay length measurement. This question
is under the study. Otherwise the time integrated
formulae should be used involving dilutions.

   The first estimate of the \Bzerod reconstruction efficiency was done
using a fast detector simulation \cite{LoI}, which simulates the response of
the detectors by smearing momenta, energies and particle identification
according to the performance of each detector. Following cuts were
used:

\begin{itemize}

\item
  all four final-state particles are within the tracking volume;
\item
  the \Kzeros decay length in the transverse plane is
  between $1~cm$ and $30~cm$;
\item
  the angle between the \Kzeros and the \Jpsi $<45^{o}$.

\end{itemize}

Muon identification
efficiency was assumed to be $80\%$ and track-finding efficiency to be $95\%$.
The reconstruction efficiency for rapidity coverage $|\eta|<2.5$  was
$\epsilon=0.13$. Expected rate of reconstructed events in one year of
running at \lumin is 1490 . The muon trigger considered
here is with $p_{t}>  20~GeV$. The dilution from wrong muon
tags within this trigger threshold was found to be
$$ D={N(correct tags)-N(mistags) \over N(correct tags)+N(mistags)} =
0.77.$$
Using the  time dependent
analysis  the error of 0.14 on CP-violation parameter measurement
could be reached within 2 years
of running.

  One of the ways of increasing statistics is to extract the CP-
violation parameter from decay of \BJKs , \KKP , \PGG .
 The branching ratio, following ARGUS and CLEO results
\cite{LPHep} is around the same as for \Jpsi \Kzero channel. This possibility
of measurement by ATLAS
is under the study. In the case that \Jpsi \Kzeros states will not be
recognized ( by an extra \photon -pair), they will be a source of
background to the studied reaction
(Figure).
The amount of this background strongly
depends on resolution.

  More detailed simulation of the detector response should be done
taking also into account background muons and \piplus, \pimin
from the same \bbarb event.

{\bf FIGURE: Effective mass spectrum of (\Jpsi \Kzero) system
      originated from reaction: \BJK (solid line), \BJKs (dashed line),\\
      \BJKP (dotted line).}

\section*{The Extraction of the \Bzero decay time from the event}

The decay time of \Bzero will be obtained by
a two-step process. First step consists of the event reconstruction
(the correct identification of all members of the \Bzero decay chain) and
the second one gives precise tracking of all particles, which ends
in the $\tau_{\Bzero}$ value.

The procedure relays on the muon detectors, which enables reconstruction of
all \muon and thus also \Jpsi. The moment of \Bzero is obtainable in the
first approximation by {\em Fox-Wollfram Moments}. Innermost silicon tracker
({\em SITV}) enables reconstructing of the second vertex and the distance
of the first and second vertex gives proper decay time of \Bzero. With
mentioned geometrical and kinematical constrains, all other members of
the decay-chain can be found, which further improves the reconstruction of
the event and consequently the $\tau_{\Bzero}$.

\section*{Conclusion}

   The preliminary results are encouraging and more detailed simulations
are under way.

\end{document}